\documentclass[aps,pra,reprint,groupedaddress,showpacs]{revtex4-1}

\usepackage{amssymb}
\usepackage{amsfonts}

\usepackage{graphicx}% Include figure files
\usepackage{dcolumn}% Align table columns on decimal point
\usepackage{bm}% bold math

\usepackage{amsmath}
\usepackage{amssymb}
\usepackage{amsbsy}
\usepackage{amsthm}
\usepackage{amsfonts}
\usepackage{mathrsfs}
\usepackage{bm}                                     	
\usepackage{mathrsfs}
\usepackage{graphicx}

\newcommand{\lv}{\left \vert}

\newcommand{\ra}{\right \rangle}
\newcommand{\ket}[1]{\lv #1 \ra}

\newcommand{\tr}{\mathrm{Tr}}

\makeatletter

\newcommand{\Rmnum}[1]{\expandafter\@slowromancap\romannumeral #1@}
\makeatother
\begin{document}

%\bibliographystyle{revtex}
% Use the \preprint command to place your local institutional report
% number in the upper righthand corner of the title page in preprint mode.
% Multiple \preprint commands are allowed.
% Use the 'preprintnumbers' class option to override journal defaults
% to display numbers if necessary
%\preprint{}

%Title of paper
\title{Measurement-based Quantum Computation Under Different Types of Noises}

% repeat the \author .. \affiliation  etc. as needed
% \email, \thanks, \homepage, \altaffiliation all apply to the current
% author. Explanatory text should go in the []'s, actual e-mail
% address or url should go in the {}'s for \email and \homepage.
% Please use the appropriate macro foreach each type of information

% \affiliation command applies to all authors since the last
% \affiliation command. The \affiliation command should follow the
% other information
% \affiliation can be followed by \email, \homepage, \thanks as well.
\author{Ding Zhong}
\author{Jian Wang}
\author{Ning Dai}
\author{Liang-Zhu Mu}
\email{muliangzhu@pku.edu.cn}
\affiliation{School of Physics, Peking University, Beijing 100871, China}

\author{Heng Fan}
\email{hfan@iphy.ac.cn}
\affiliation{Institute of Physics, Chinese Academy of Science, Beijing 100190, China}

%Collaboration name if desired (requires use of superscriptaddress
%option in \documentclass). \noaffiliation is required (may also be
%used with the \author command).
%\collaboration can be followed by \email, \homepage, \thanks as well.
%\collaboration{}
%\noaffiliation

\date{\today}

\begin{abstract}
Measurement based quantum computation (MBQC) is an effective paradigm for universal quantum computation.
In this scheme, the universal set of quantum gates are realized
by only local measurements on the prior prepared cluster states.
The inevitable decoherence is harmful to the realization of those quantum gates.
Here, we investigate the performance of the quantum gates exposed to different type of noises.
We find that some errors may not influence the success of the quantum gates, in contrast, some others
may destroy their realization.
We show that there is a controlling pattern that can protect quantum gates from certain types of noises
and thus can improve the success probability of the gates implementation.
\end{abstract}

% insert suggested PACS numbers in braces on next line
% insert suggested keywords - APS authors don't need to do this
%\keywords{}

%\maketitle must follow title, authors, abstract, \pacs, and \keywords
\maketitle

% body of paper here - Use proper section commands
% References should be done using the \cite, \ref, and \label commands
\section{Introduction}

Quantum computers are more powerful than their classical counterpart,
as displayed in solving factoring problems by Shor algorithm \cite{shor1999polynomial}. Since then, much focus has been paid on the realization of quantum computation
and quantum information processing. Measurement-based quantum computation (MBQC) is a paradigm for universal quantum computation \cite{PhysRevLett.86.5188}.
By adaptive local measurements on a highly entangled states named as cluster states, the universal sets of quantum gates
can be realized. By this scheme, the difficulty in realizing the quantum gates is simplified
as only performing the much easier local measurements, while the resource of cluster states are assumed to be available.
Then the preparation of various necessary cluster states, which are general multiparticle
entanglement, becomes crucial.
It has been shown that the structure of such cluster states can be understood as a
valence-bond solid with only nearest neighbor bonds \cite{PhysRevA.70.060302}, and
thus the universal quantum computation is realized.
It is also known that the MBQC essentially works with the help of quantum teleportation
\cite{PhysRevA.71.032318,PhysRevA.68.022312,gottesman1999demonstrating}.

Many efforts have been made to construct the desired many-body entangled states that are appropriate for quantum computation.
For some specified systems,
the ground state of the Hamiltonian is found to be universal for quantum computation \cite{PhysRevA.71.062313,PhysRevLett.110.120502,jianming}.
In particular, valence-bond Hamiltonians, which enjoys advantage of two-body interaction, is shown to be universal \cite{PhysRevA.70.060302,PhysRevA.74.040302,PhysRevLett.101.010502,PhysRevLett.102.220501,PhysRevA.85.010304,PhysRevLett.106.070501,miyake2011quantum}.
The preparation of cluster states in optical system has been experimentally achieved\cite{walther,knill,nielsen,gao}.

The multipartite entangled state is in general fragile because of decoherence.
So the protection of entanglement of cluster states
plays a central role for MBQC in increasing the accuracy and the longevity of calculation.
If we consider that the cluster state is simply the ground state of a corresponding Hamiltonian.
The decoherence induced by finite temperature should be the main source of error.
Recently, it is shown that
the thermal state of an interacting cluster Hamiltonian exhibits less decoherence compared with
the non-interacting Hamiltonian at equal temperature \cite{PhysRevLett.110.120502}.
Realistically, the cluster state might be directly prepared but not by cooling the system to zero temperature
for creating the ground state of the Hamiltonian. In this sense, the cluster states are
exposed to various noises.
The sources of decoherence are not limited to just thermal noise, but may also include interactions with environment\cite{bellomo,chaves,santos,arruda,daniel,roman}, experimental deviation, etc. Consequently, we should take variety of decoherence under consideration.

Rather than focusing on making it less error prone, the influence of different types of decoherence that occurs on different qubits of cluster state
upon the quantum computation still remains unknown. Plainly, the original cluster states are usually influenced by a variety of possible types of noise.
We are interested in evaluating how much damage exactly that each type of noise bring about to the final resource state,
whether the noise on different qubit of the original cluster state has the same impact on the final implemented gate.
If this question is clarified, one can focus on diminishing the noise of the most harmful type on most crucial qubit by means of error correction
or other possible techniques. This will save the precious quantum information resource.
In addition, we also investigate whether there is a pattern to control the qubits that would best improve the fidelity of quantum gates.

Our article is organized as following: In the next section, we consider the situation that the error occurs on a single qubit of the cluster states.
We will run all all qubits of the cluster states for different quantum gates.
In section III, we try to investigate whether there are patterns to control particular qubits on the lattice that
keep fidelity at a higher level. In the last section, we will present a brief conclusion.

\section{Decoherence on a single qubit}

First, let us introduce the notations. The pauli matrices labeled $X$,$Y$,$Z$ are given by:
\begin{eqnarray}
X=\left(
\begin{array}{cc}
0&1\\
1&0\\
\end{array}
\right),
Y=\left(
\begin{array}{cc}
0&-i\\
i&0\\
\end{array}
\right),
Z=\left(
\begin{array}{cc}
1&0\\
0&-1\\
\end{array}
\right).
%
%$$
%X=\begin{pmatrix}
%0&1\\
%1&0\\
%\end{pmatrix}
%
\end{eqnarray}
The necessary cluster states for universal set of gates are represented in FIG. 1. We generally
use the same notations are those used in \cite{PhysRevLett.110.120502}.
The cluster stabilizer on a lattice $\mathcal{T}$ is defined by $K_i=X_i\bigotimes_{j\in N_i}Z_j$ for each site i, where $N_i$ denotes the set of vertices that are adjecent to the site i. The cluster state is defined by simultaneous eigenstate of all cluster stablizer $K_i$ with eigenvalue $+ 1$.

\begin{figure}
\begin{center}
\includegraphics[width=40mm]{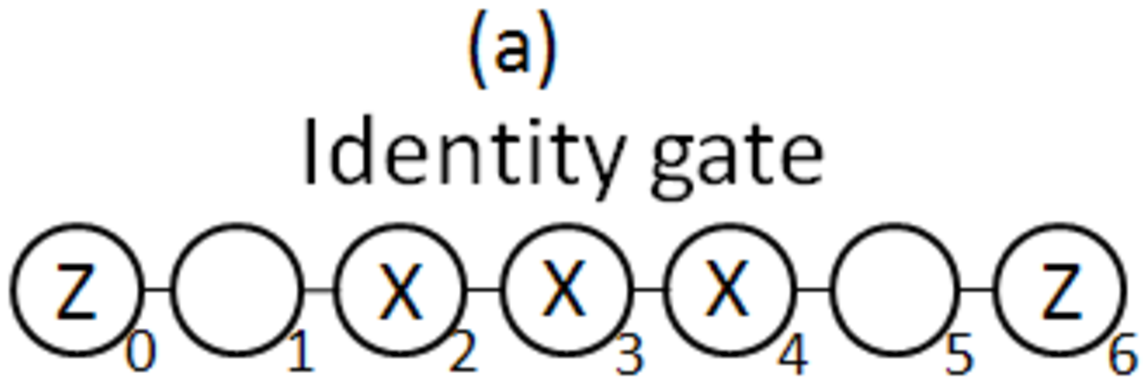}
\includegraphics[width=40mm]{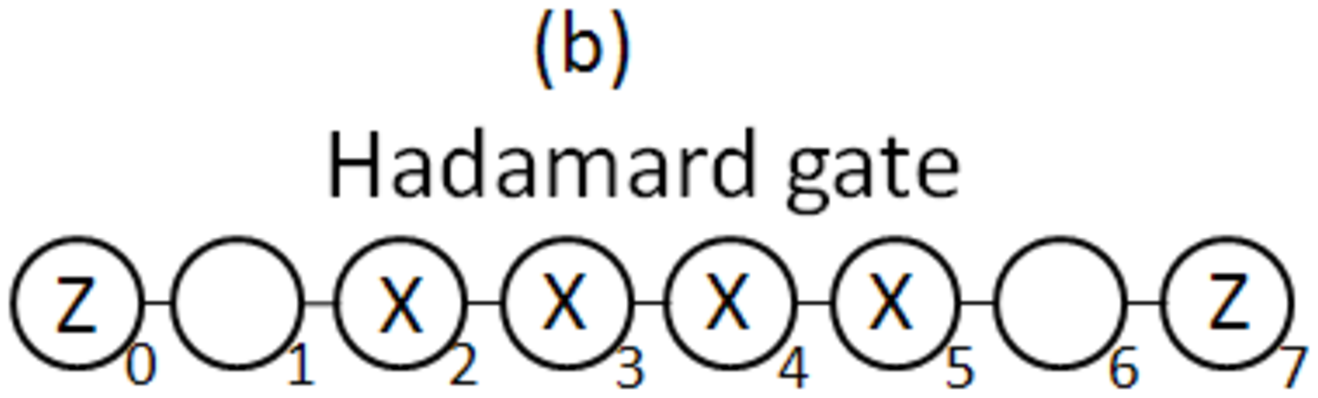}
\includegraphics[width=40mm]{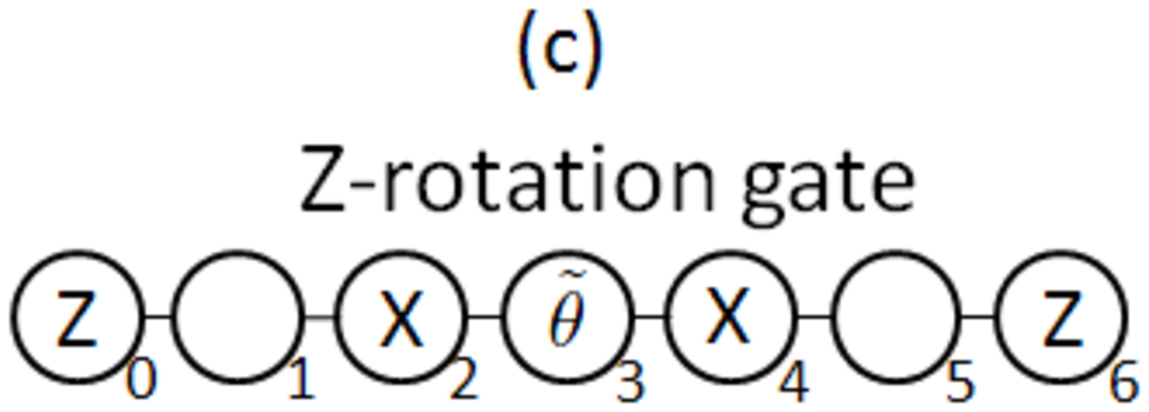}
\includegraphics[width=40mm]{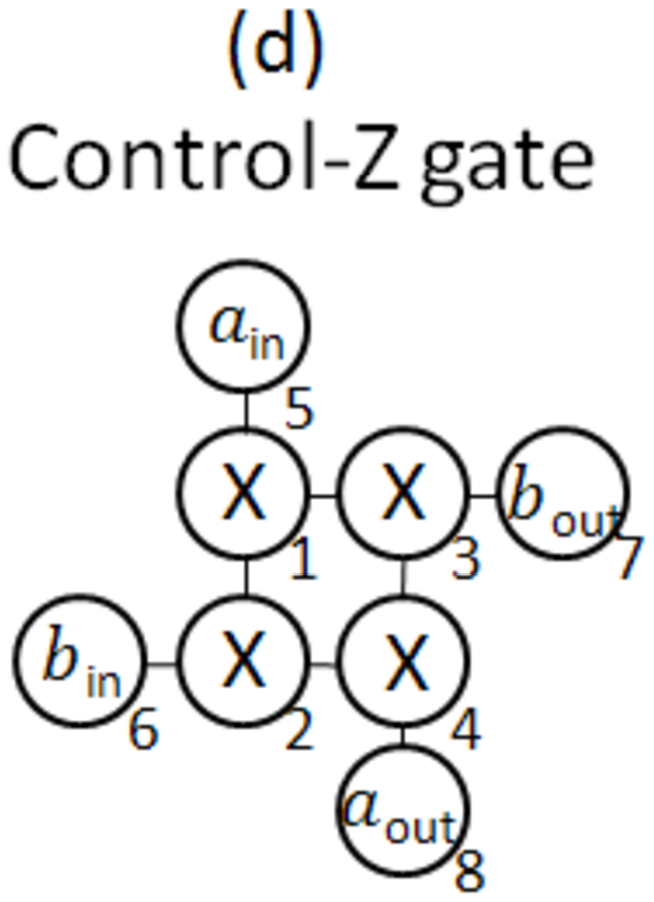}
\caption{Universal set of quantum gates. The measurement patterns are shown in the figure. (a) the identity gate. (b) the Hadamard gate.
(c) the Z-rotation gate with the angle $\theta$. In this figure, the mark $\tilde{\theta}$ denotes the qubit is measured in the basis of $\cos{\tilde{\theta}}X+\sin{\tilde{\theta}}Y$ where $\tilde{\theta}=m_2\theta$, and $m_2$ represents the measurement result of qubit 2.
(d) the controlled-Z gate. \label{measurementpattern}}
\end{center}
\end{figure}

The group that generated by multiplication of Pauli matrices is known as Pauli group. The Clifford group is defined as the group of unitary operators that map the Pauli group to itself, which is for teleportation-based quantum computation. A universal set of gates can be identity gate, Hadamard gate, Z-rotation gate,
and controlled-Z gate. These four gates are called universal because the combination of them is enough to perform any quantum computation. To prepare such four gates on MBQC scheme, the measurement sequence are shown in fig\ref{measurementpattern}, and each qubit is labeled with a number. After the local measurement, a correction unitary is performed which is dependent on measurement result of local measurement.

On purpose is to evaluate the performance of the prepared gates by fidelity.
The gate fidelity between the source state that is prepared and the ideal gate can be calculated
by:
\begin{eqnarray}
F(\rho)=\tr{\left(\rho\rho_{ideal}\right)},
\end{eqnarray}
where $\rho $ is the final state with noise, and $\rho_{ideal}$ is the state for noiseless condition.
Since the gate prepared is dependent on the measurement result, the fidelity is dependent on the measurement result consequently (only for the case that the original state is ideally cluster state, the gate prepared out of it has a fidelity fixed at 1 which is independent on the measurement result).
Fortunately, it has been shown that the average correlation function weighted by probability can be calcucalated by several concise formula \cite{chung2009characterizing}. With some calculations, we can obtain the expectation value of fidelity:

\begin{widetext}
\begin{align}
F_{\rm id}(\rho) &= \tr \biggl( \rho \frac{1+\prod_{k=1}^{3} K_{2k-1}  }{2} \frac{1+\prod_{k=1}^2 K_{2k}}{2} \biggr), \label{a1} \\
F_{H}(\rho) &= \tr \biggl( \rho \frac{1+\prod_{k=1}^{3} K_{2k-1}  }{2} \frac{1+\prod_{k=1}^{3} K_{2k}}{2} \biggr), \label{a2} \\
F_{U_Z(\theta)} (\rho) &= \tr \biggl( \rho \frac{1+K_2 K_4 }{2} \frac{1+K_1 K_3 K_5 (\cos^2 \theta + \sin^2 \theta K_4) + \cos \theta \sin \theta (Z_0 Y_1 Z_2) K_2 K_3 (1-K_4) K_5}{2} \biggr), \label{a3} \\
F_{CZ} (\rho) &= \tr \biggl( \rho \frac{I+ K_{a_{\rm in}} K_3 K_{a_{\rm out}} }{2} \frac{I+ K_{b_{\rm in}} K_4 K_{b_{\rm out}}}{2} \frac{I+ K_1 K_4}{2} \frac{I+ K_2 K_3}{2}  \biggr). \label{a4}
\end{align}
\end{widetext}
As we mentioned, here, $\rho$ denotes the cluster state affected by a noisy quantum channel before the measurement. These formulas illustrate that despite of the uncertainty of measurement result, the average fidelity can be shown prior to measurement, right after the cluster state is prepared. This idea is inspiring and useful because it tells us that the algorithm which takes exponential complexity to cover all possible measurement result is not necessary and can be simplified to several neat formulas.

In our model, we assume that the error happens individually on qubits. This is a reasonable assumption,
and thermal state is a typical example for this case. The state $\rho$ with independent and individual noise satisfies the form:
\begin{eqnarray}
\rho= \sum_{i_1,i_2\cdots i_l=1,\cdots n}  E_{i_1}^{(1)}E_{i_2}^{(2)}\cdots E_{i_{l}}^{(l)}\rho E_{i_l}^{(l)\dagger}\cdots E_{i_2}^{(2)\dagger}E_{i_1}^{(1)\dagger},
\end{eqnarray}
where $E_i^{(m)}$ satisfy:
\begin{eqnarray}\label{tracepreserving}
\sum_{i=1\cdots n} E_i^{(m)\dagger}E_i^{(m)}=I ~~~ \forall m,
\end{eqnarray}
Here $E_i^{(m)}$ represents operations on qubit $m$, and satisfy trace preservation condition,
$l$ is the total number of qubits. Each complete trace preserving set of $E_i^{(m)}$ correspond to one type of quantum noise.

In this article, we take four common types of noise under consideration:
\begin{itemize}
\item $X$ type of noise, the
corresponding quantum channel is represented as,
\begin{eqnarray}
E_1^{(m)}&=&
\left(
\begin{array}{cc}
\sqrt{1-p}&0\\
0&\sqrt{1-p}\\
\end{array}
\right),
\nonumber \\
E_2^{(m)}&=&\left(
\begin{array}{cc}
0&\sqrt{p}\\
\sqrt{p}&0\\
\end{array}
\right).
\end{eqnarray}

\item $Z$ type of noise, we have the representation,
\begin{eqnarray}
E_1^{(m)}&=&
\left(
\begin{array}{cc}
\sqrt{1-p}&0\\
0&\sqrt{1-p}
\end{array}
\right),
\nonumber \\
E_2^{(m)}&=&\left(
\begin{array}{cc}
\sqrt{p}&0\\
0&\sqrt{p}
\end{array}
\right).
\end{eqnarray}

\item Phase damping channel:
\begin{eqnarray}
E_1^{(m)}=
\left(
\begin{array}{cc}
1&0\\
0&\sqrt{1-p}
\end{array}
\right),
E_2^{(m)}=\left(
\begin{array}{cc}
0&0\\
0&\sqrt{p}
\end{array}
\right).
\end{eqnarray}

\item Amplitude damping channel:
\begin{eqnarray}
E_1^{(m)}=
\left(
\begin{array}{cc}
1&0\\
0&\sqrt{1-p}
\end{array}
\right),
E_2^{(m)}=\left(
\begin{array}{cc}
0&\sqrt{p}\\
0&0
\end{array}
\right)
\end{eqnarray}
\end{itemize}
Here the probability of error is $p$.

On purpose of analyzing the impact that the noise bring about to the final gate we prepare, we start with the question whether each qubit in the cluster plays the role of the same importance. With this information, we can intentionally protect a
specific qubit from decoherence and thus can improve the success probability efficiently.
This exploration would also provide a starting point for analyzing more complicated situation.

\begin{figure}
\begin{center}
\includegraphics[width=40mm]{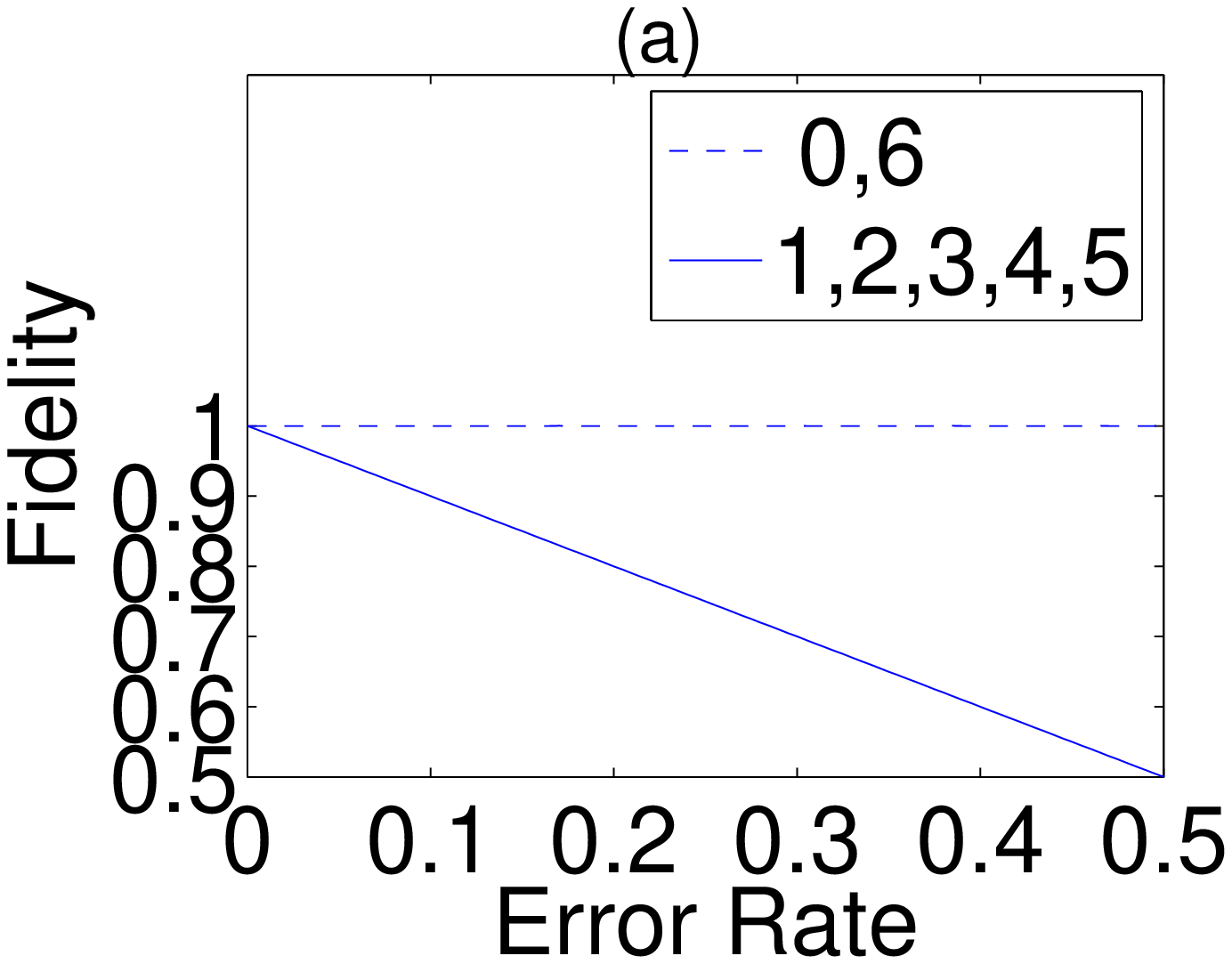}
\includegraphics[width=40mm]{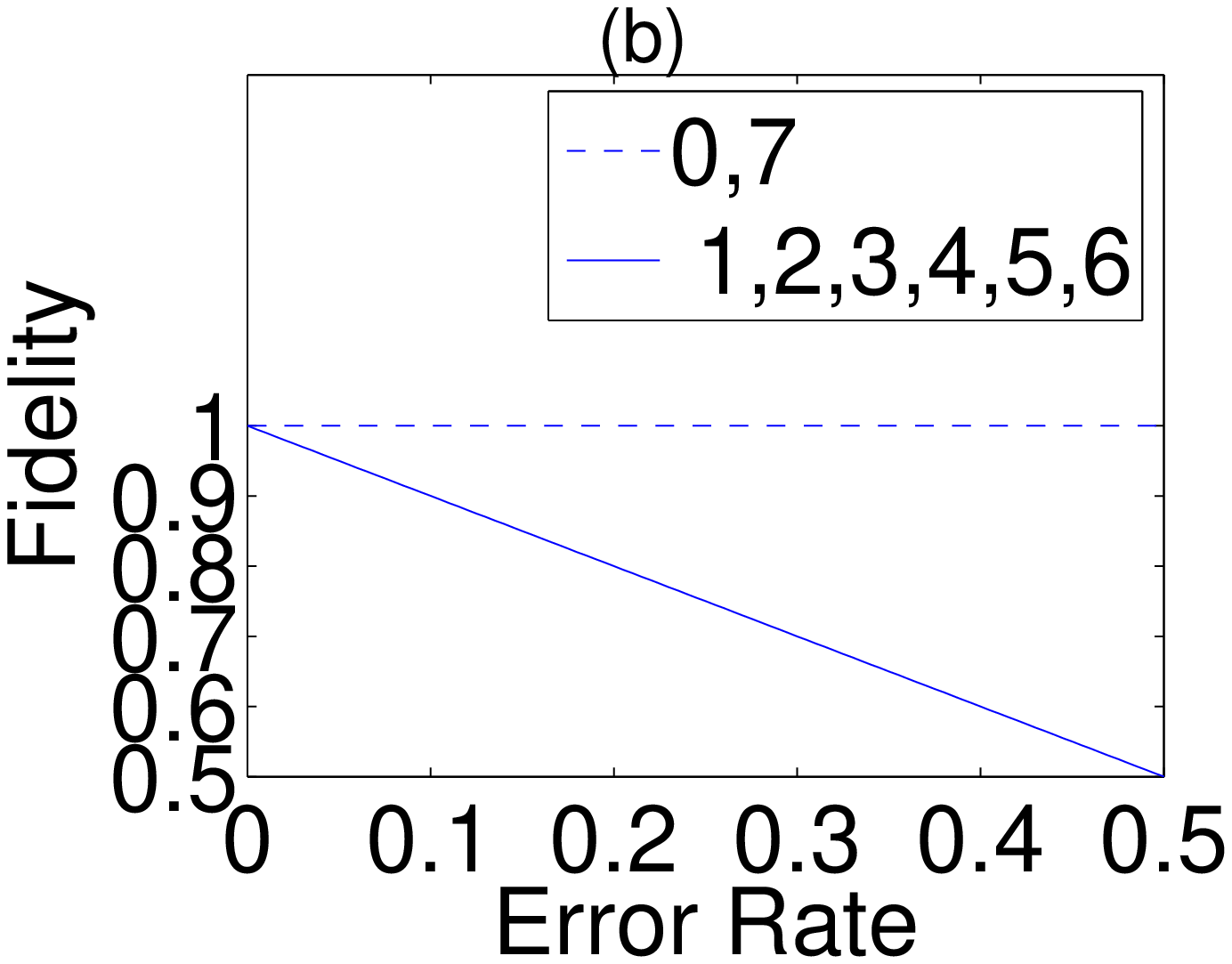}
\includegraphics[width=40mm]{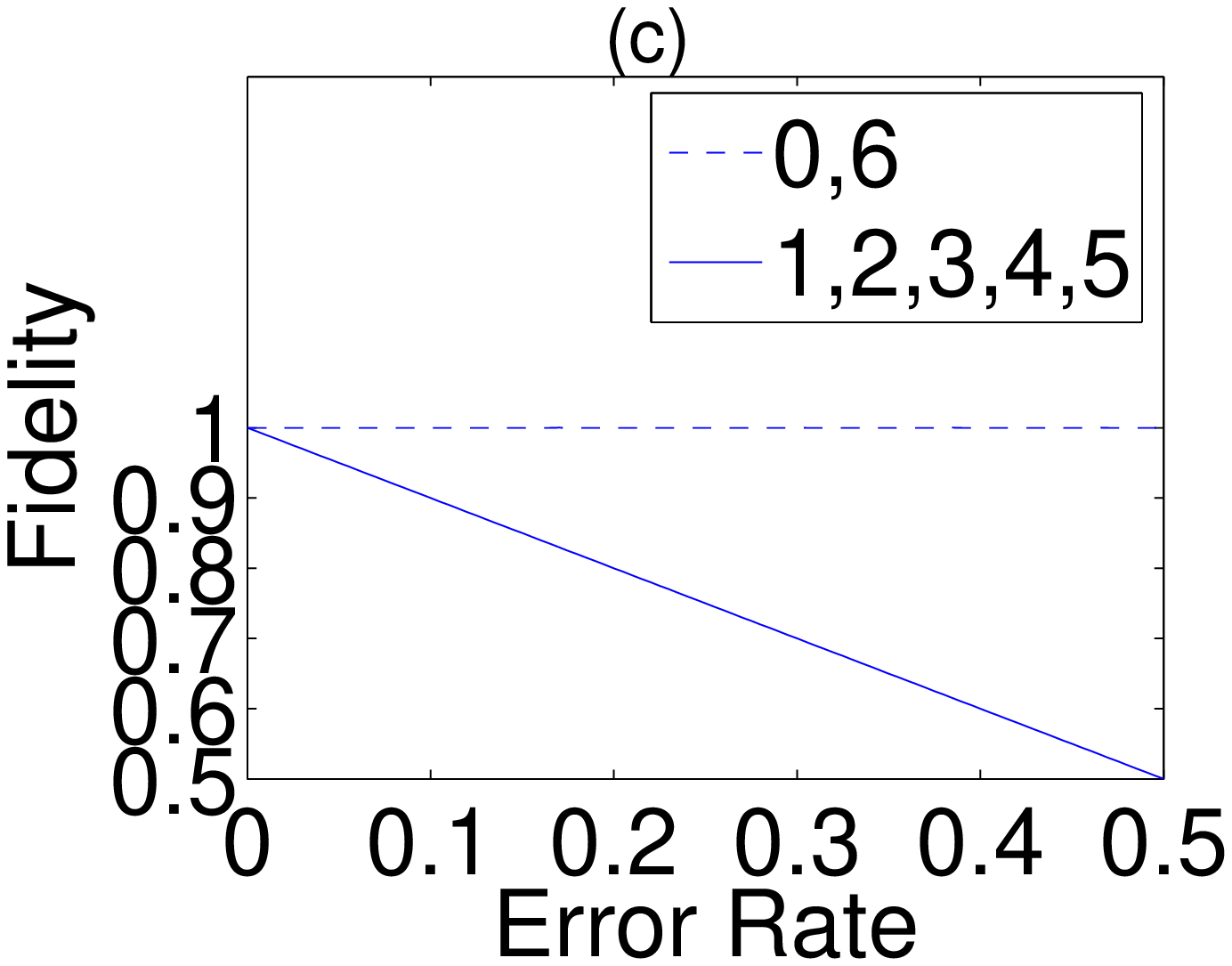}
\includegraphics[width=40mm]{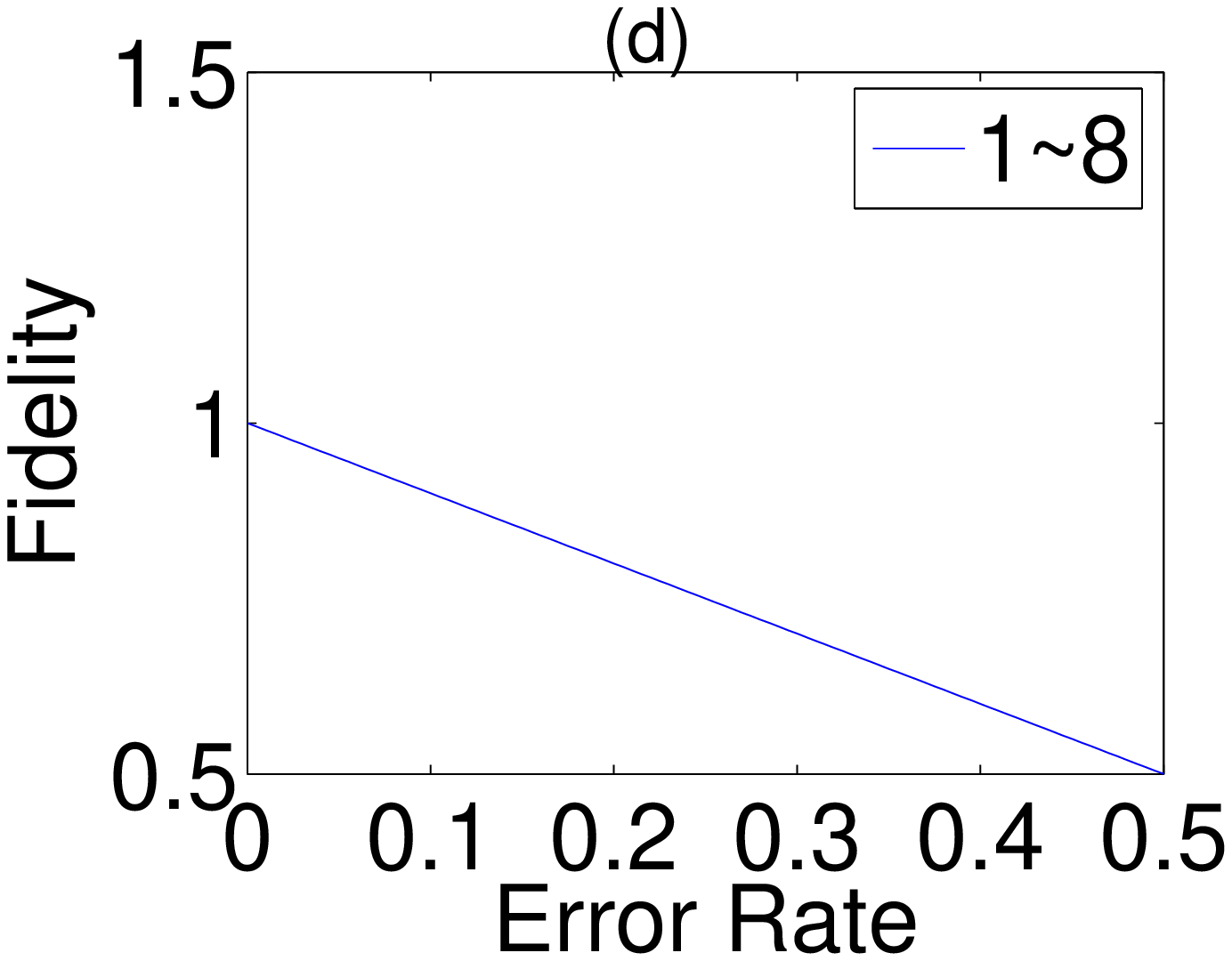}
\caption{The fidelity-error rate curve. Only one qubit is infected with Z noise, the numbers in the legend denote the qubit infected. Some of the results are the same and combined. (a)identity gate;(b)hadamard gate;(c)Zrotation gate;(d)controlled-Z gate;\label{Znoise}}
\end{center}
\end{figure}

Combining Eq. (\ref{tracepreserving}) with equations (\ref{a1})-(\ref{a4}), we can obtain the relationship between
gate fidelity with the error rate in a specific quantum channel. We show the numerical results in FIG. \ref{Znoise} to FIG \ref{Phasenoise}.
It shows the gate fidelity under four types of noises, which actually means the fidelity between
the initial cluster state and the output cluster state through those noisy quantum channels.
We conclude the results as follows.

\begin{figure}
\begin{center}
\includegraphics[width=40mm]{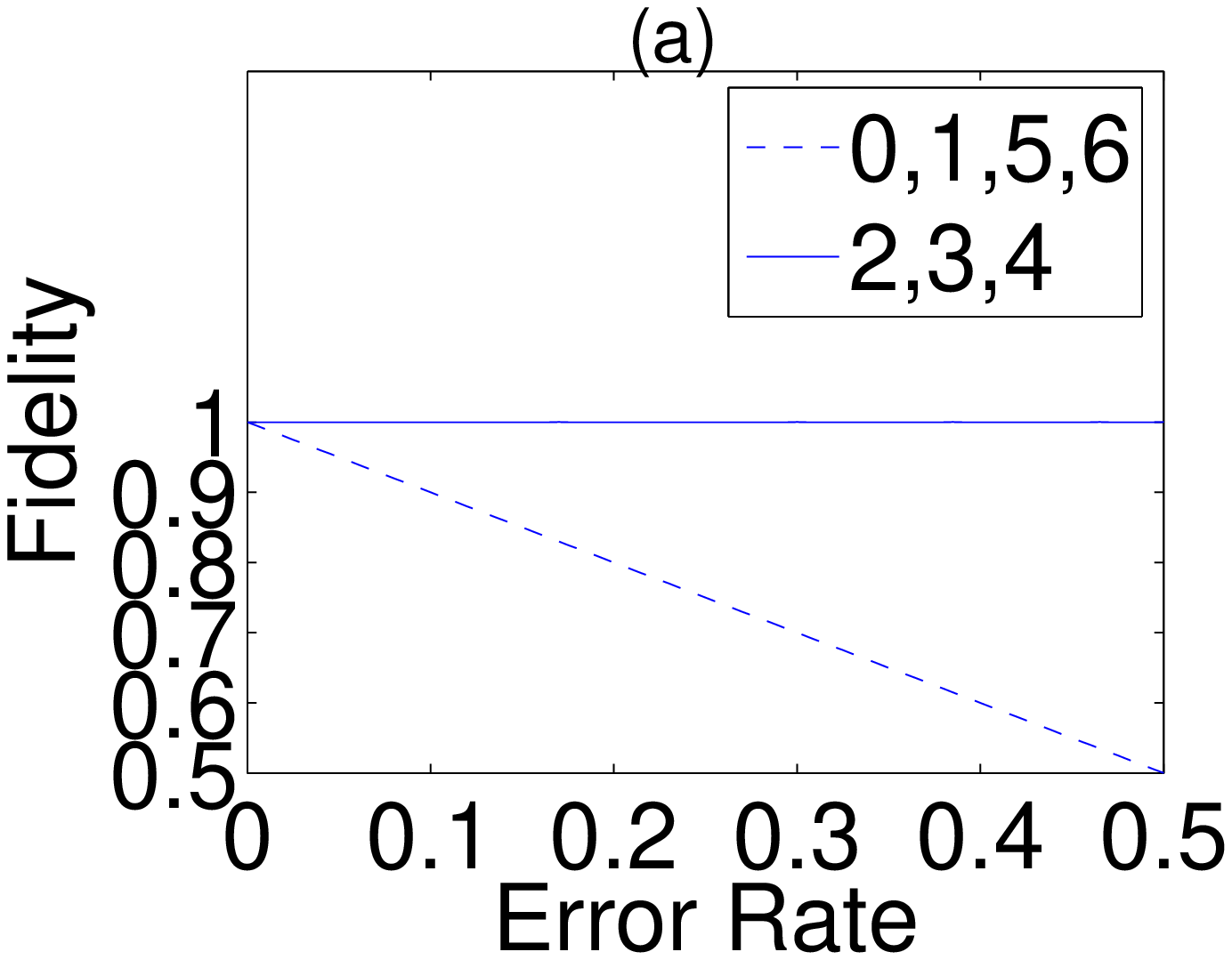}
\includegraphics[width=40mm]{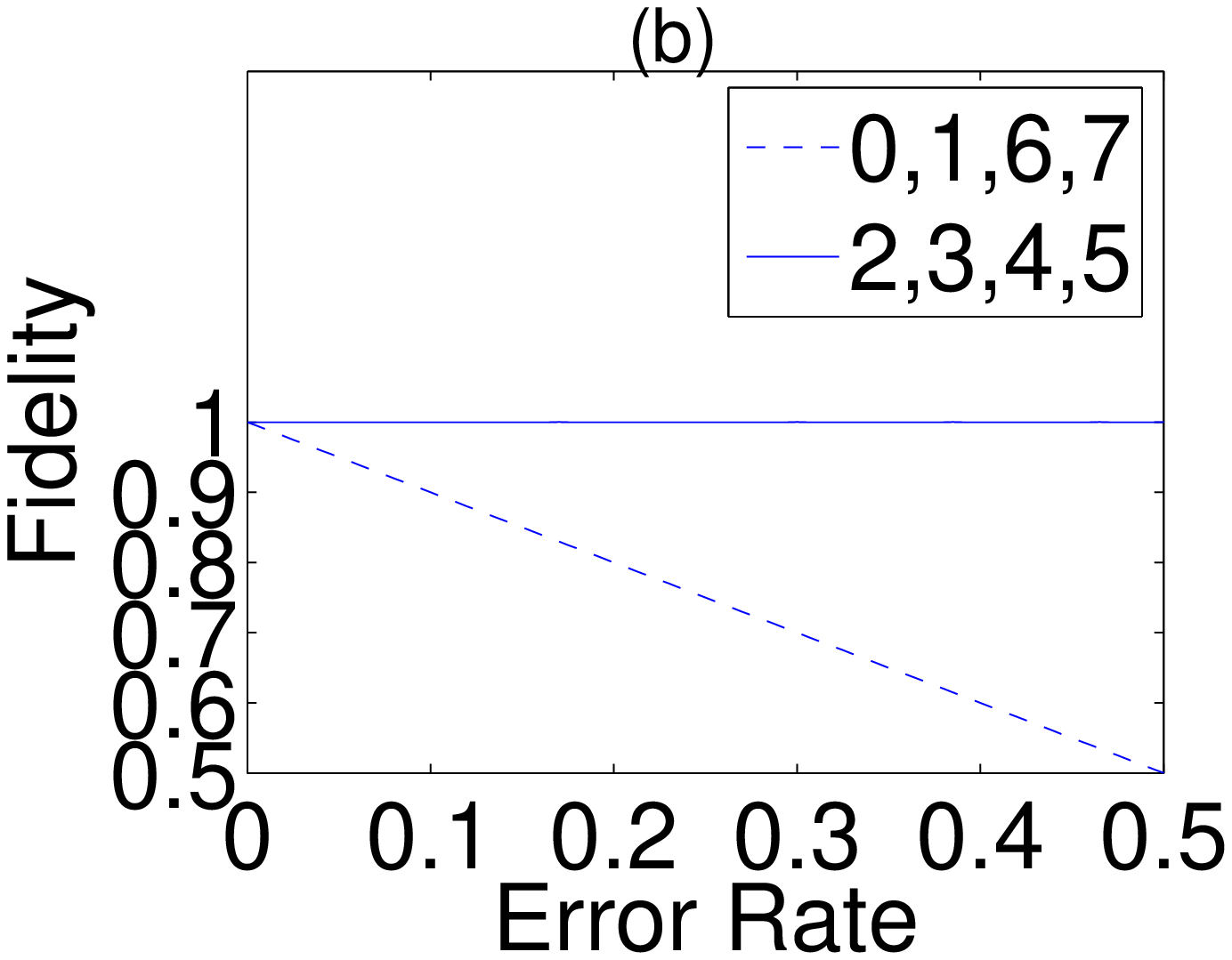}
\includegraphics[width=40mm]{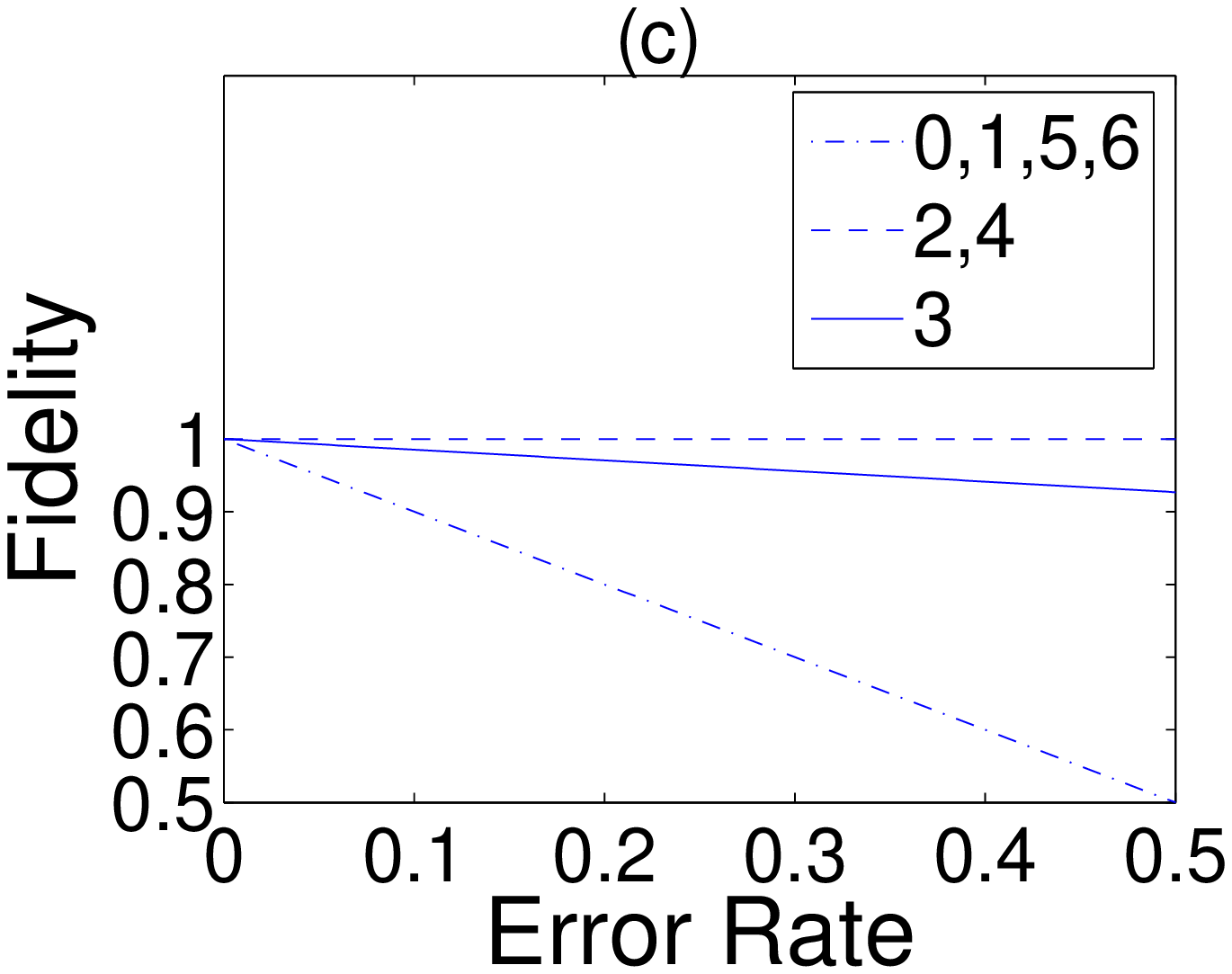}
\includegraphics[width=40mm]{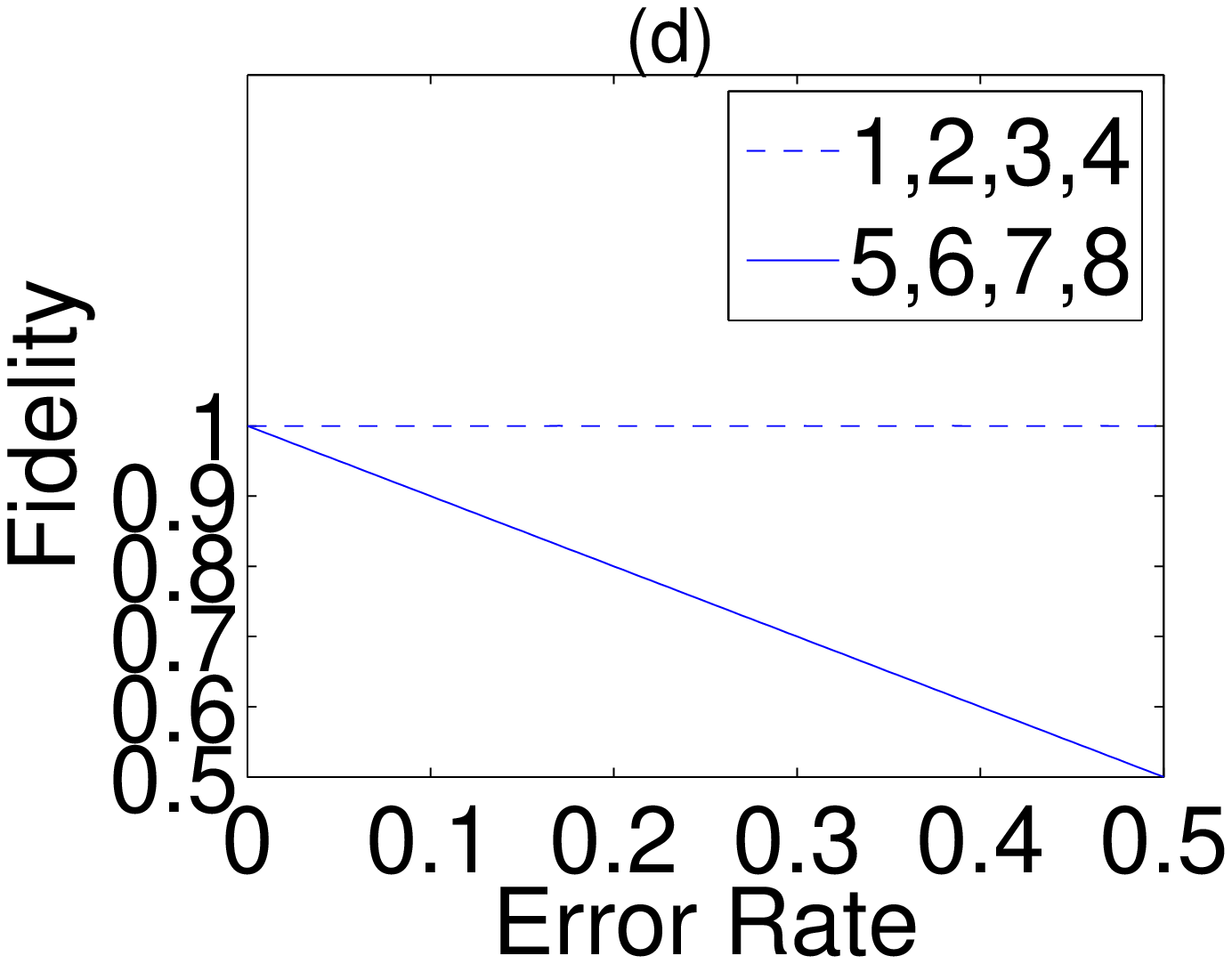}
\caption{The fidelity-error rate curve. Only one qubit is infected with X noise, the numbers in the legend denote the qubit infected. Some of the results are the same and combined. (a)identity gate;(b)hadamard gate;(c)Zrotation gate;(d)controlled-Z gate;\label{Xnoise}}
\end{center}
\end{figure}

\begin{figure}
\begin{center}
\includegraphics[width=40mm]{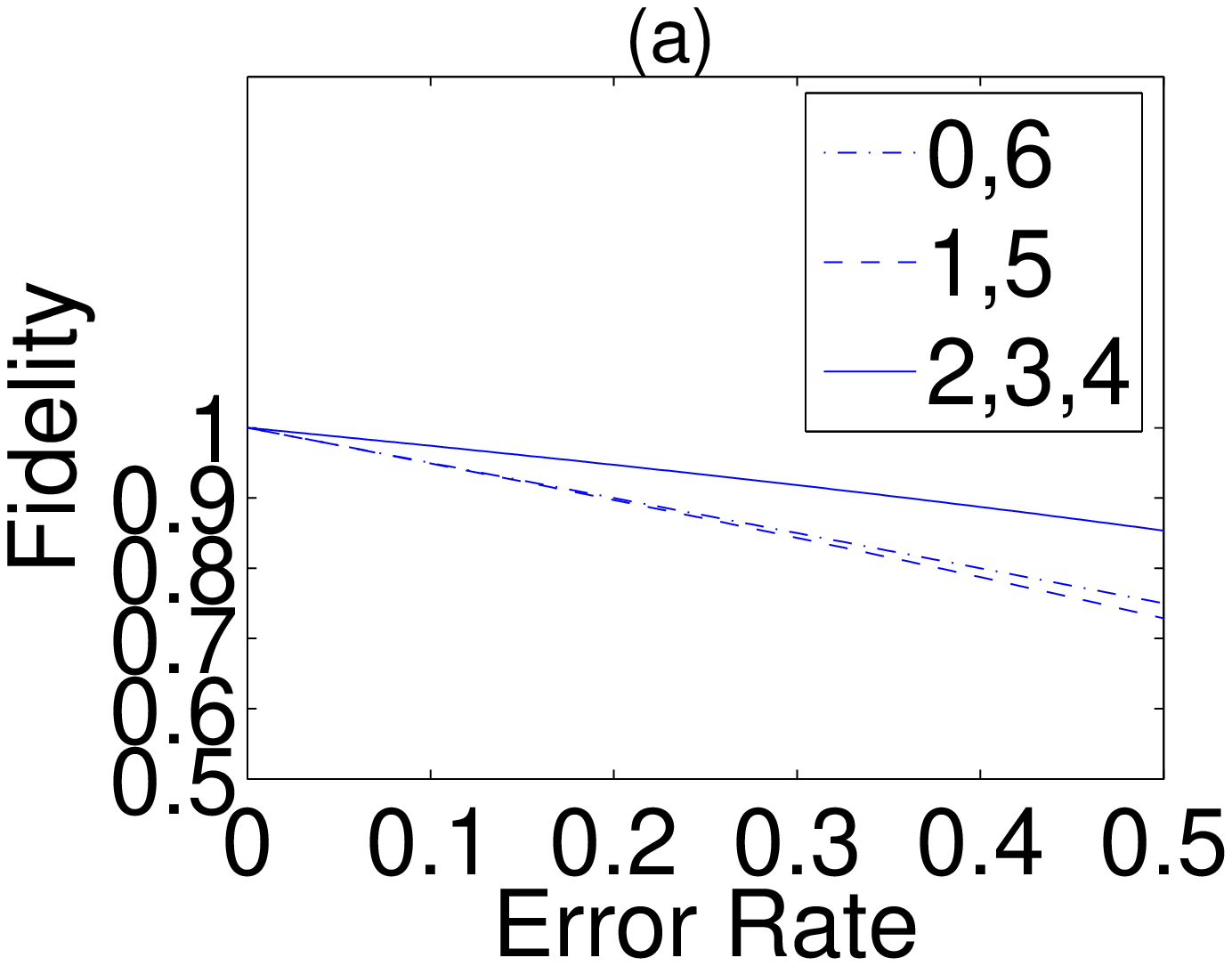}
\includegraphics[width=40mm]{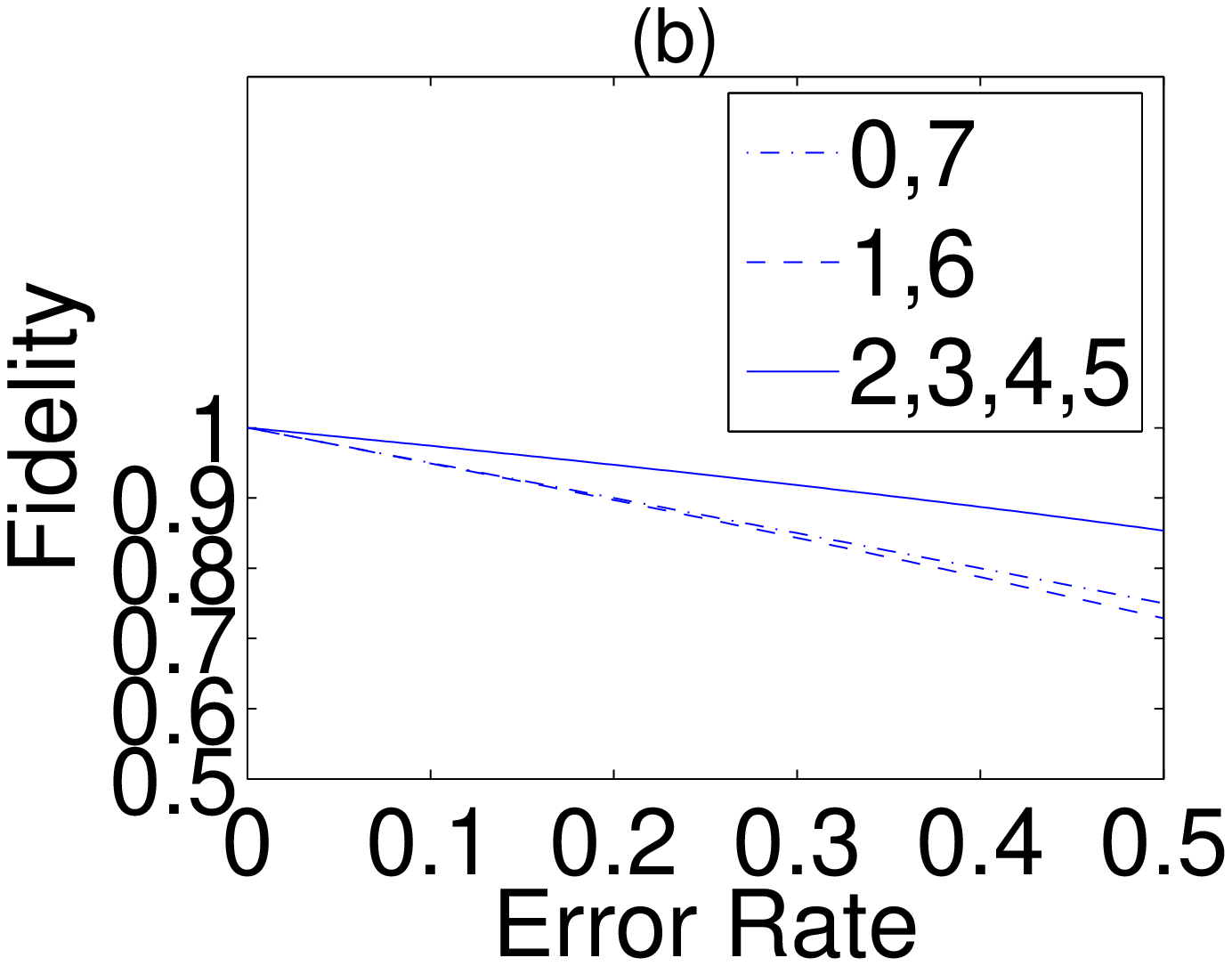}
\includegraphics[width=40mm]{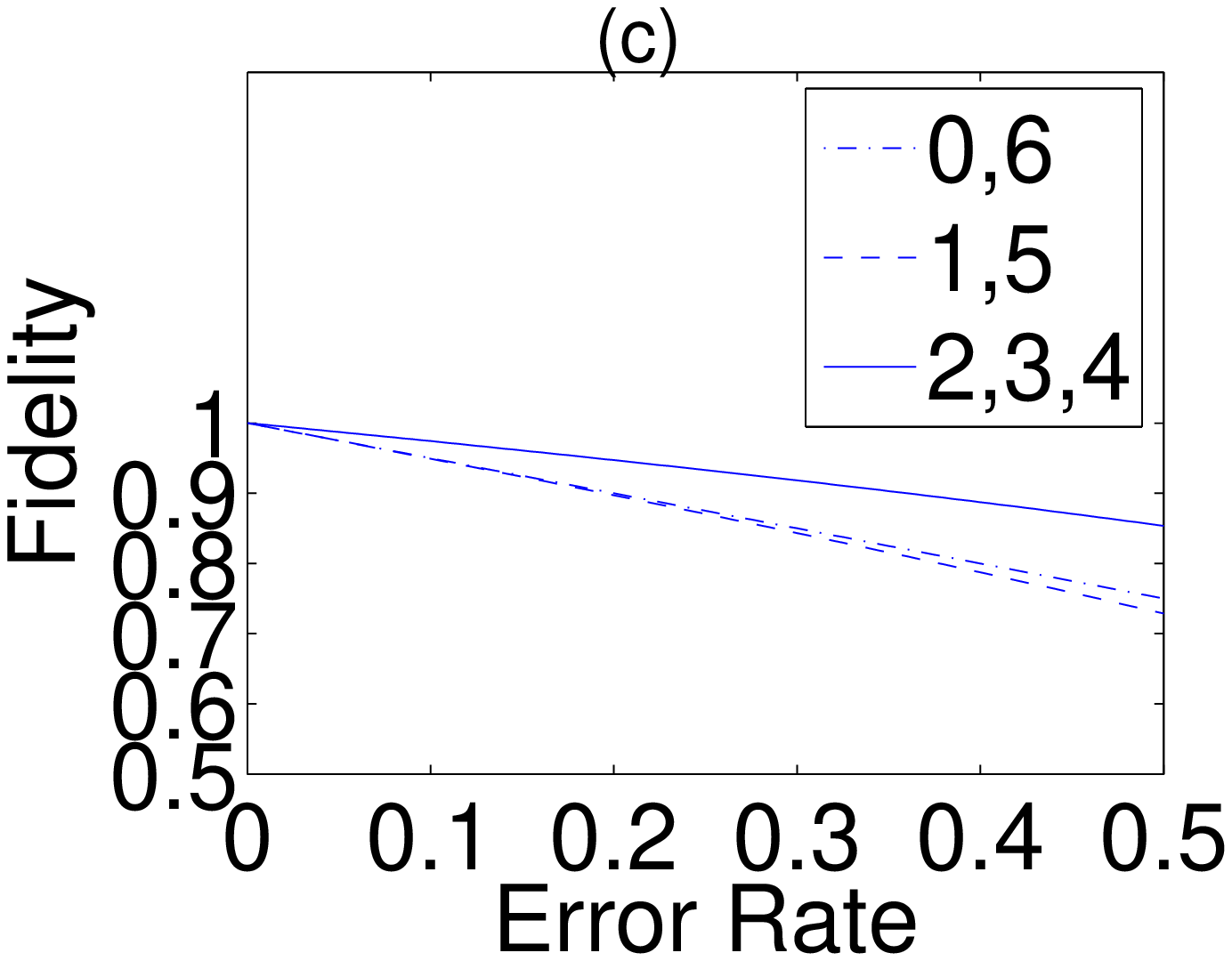}
\includegraphics[width=40mm]{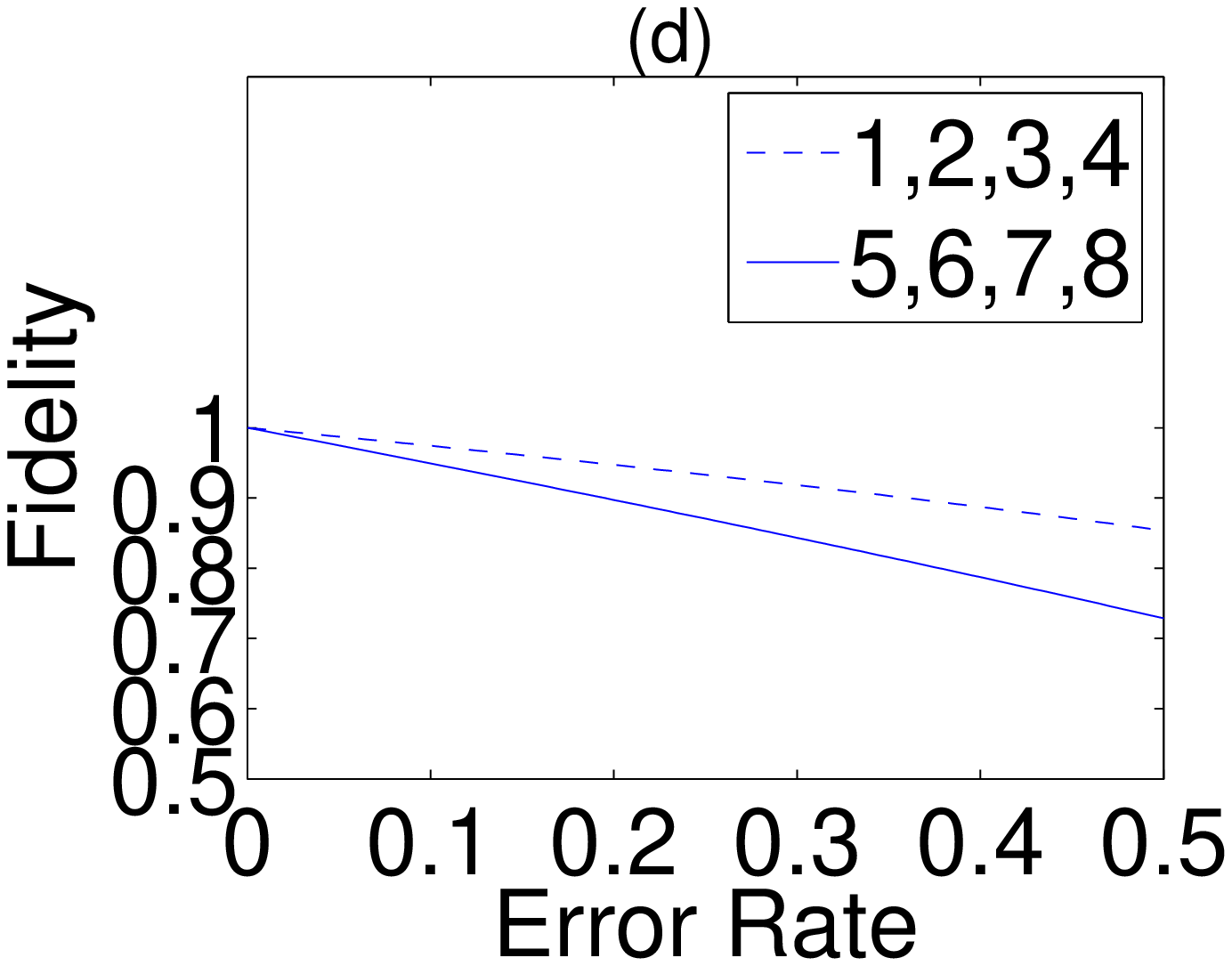}
\caption{The fidelity-error rate curve. Only one qubit is infected with Phase Dammping noise, the numbers in the legend denote the qubit infected. Some of the results are the same and combined. (a)identity gate;(b)hadamard gate;(c)Zrotation gate;(d)controlled-Z gate; \label{Ampnoise}}
\end{center}
\end{figure}

\begin{figure}
\begin{center}
\includegraphics[width=40mm]{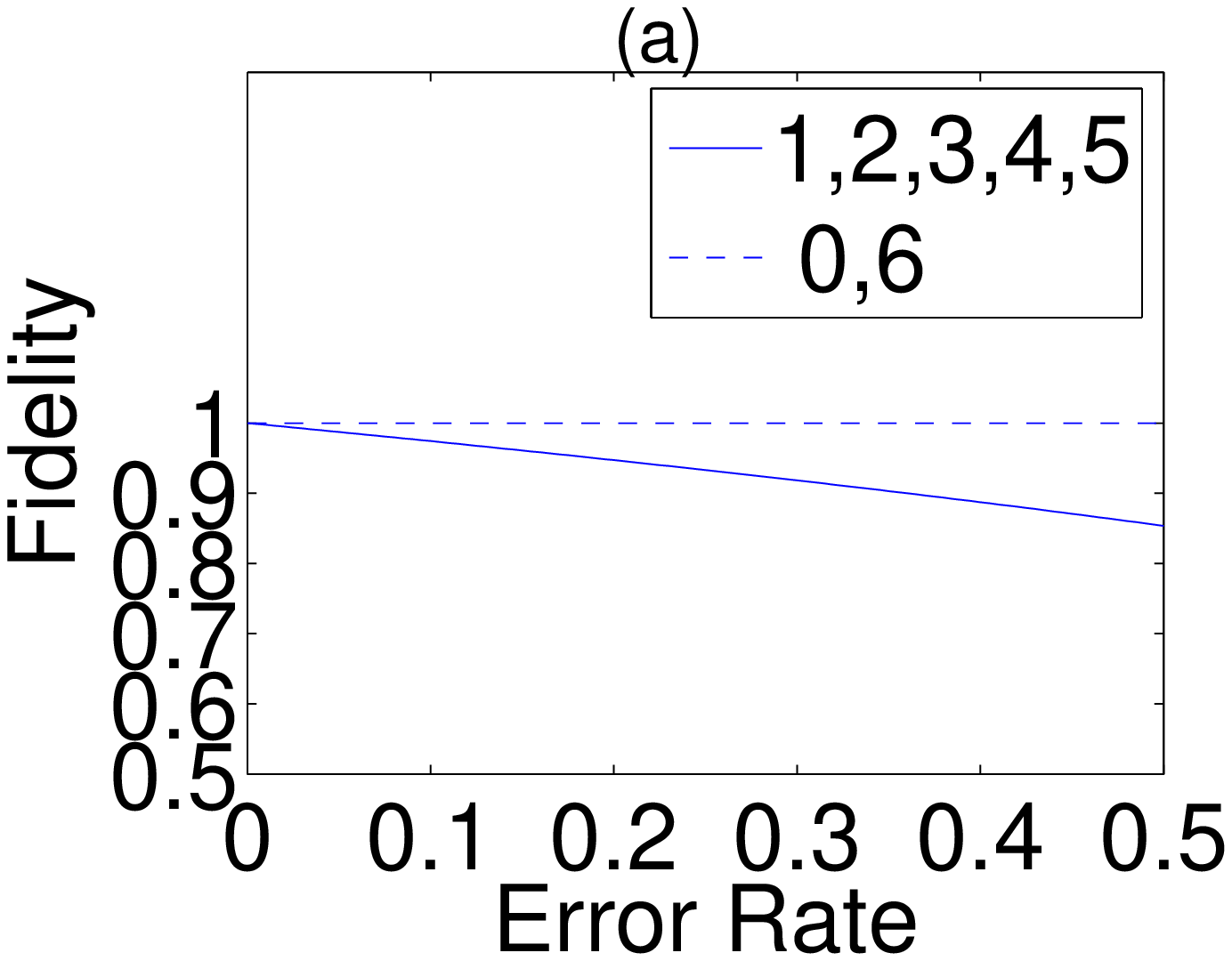}
\includegraphics[width=40mm]{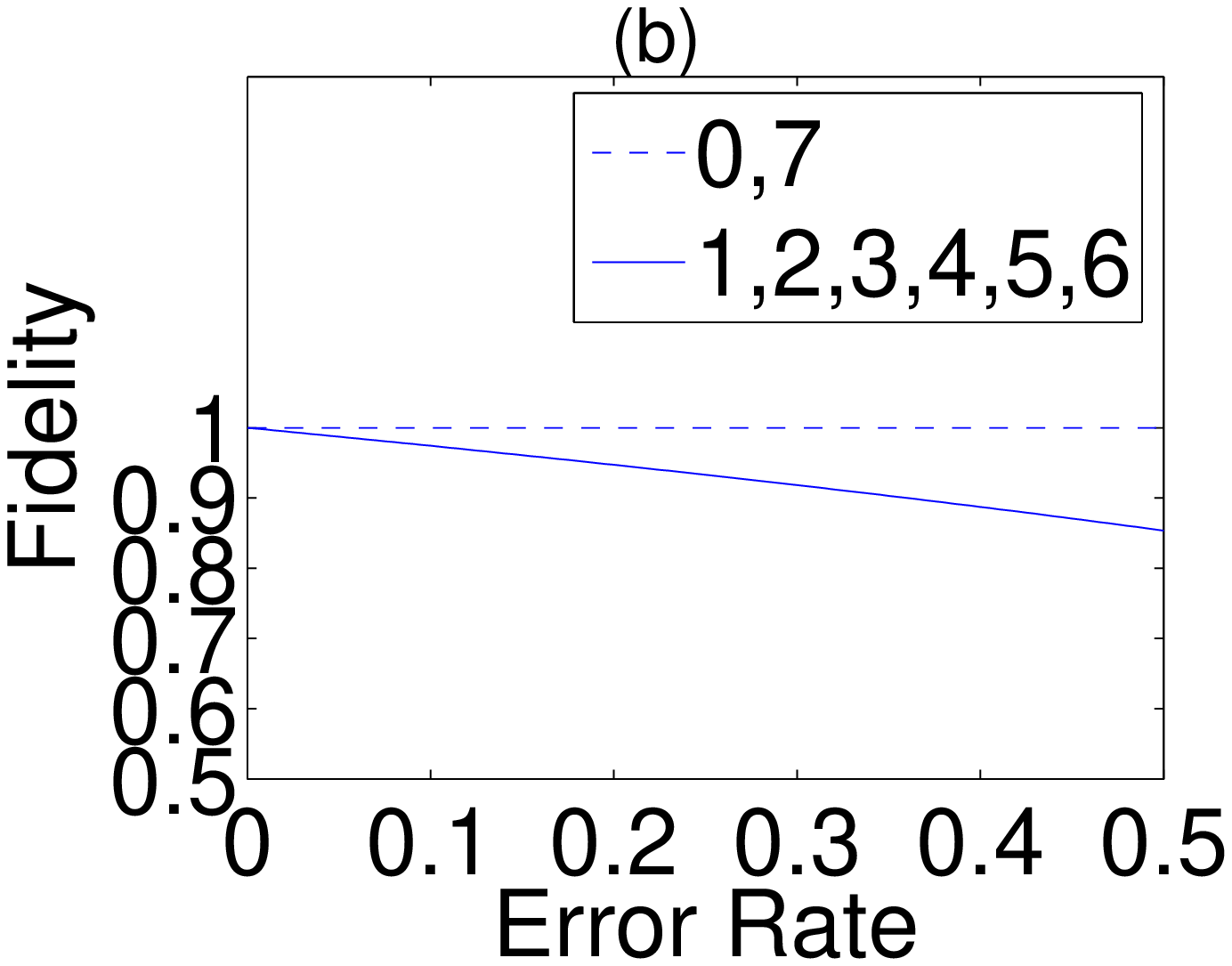}
\includegraphics[width=40mm]{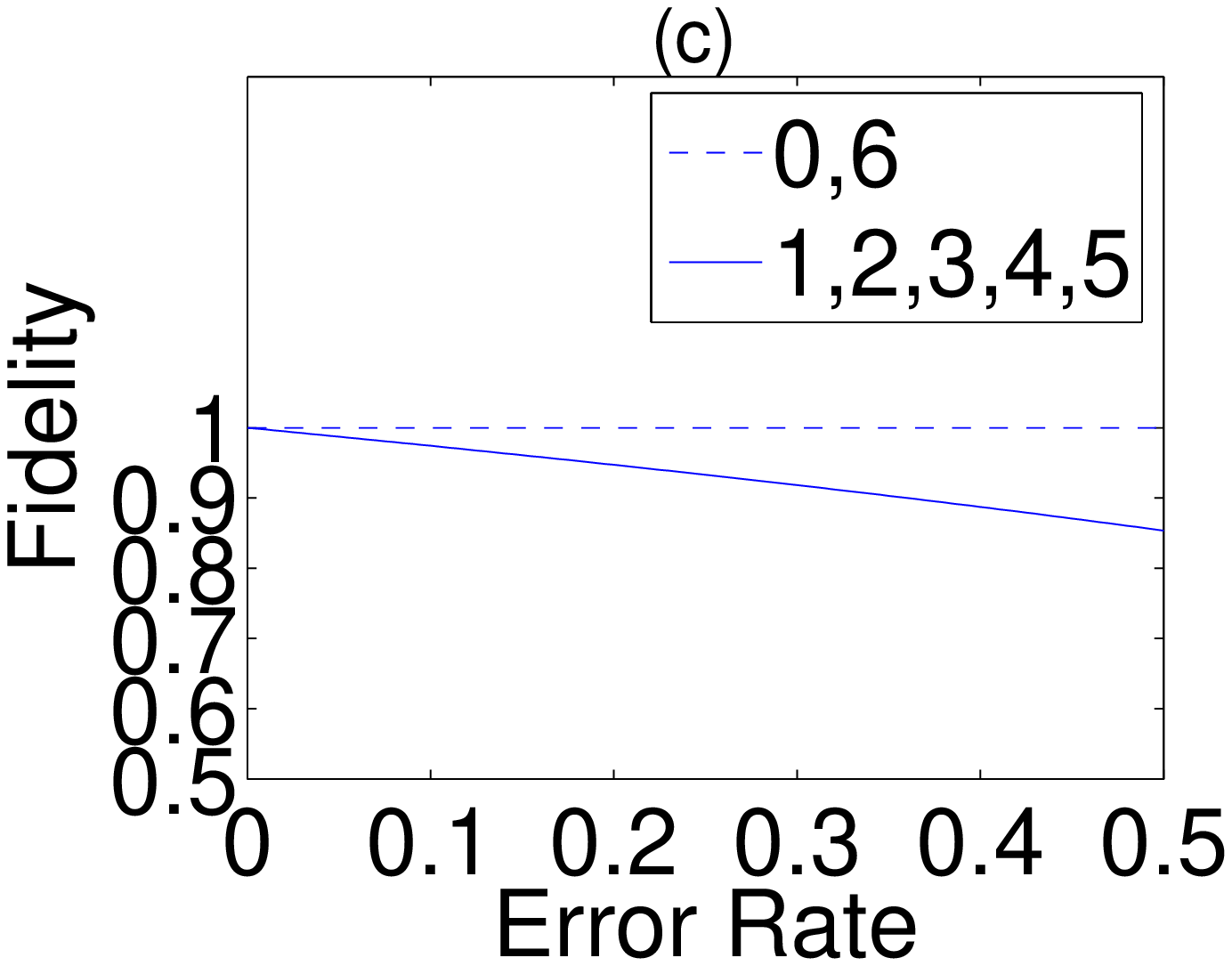}
\includegraphics[width=40mm]{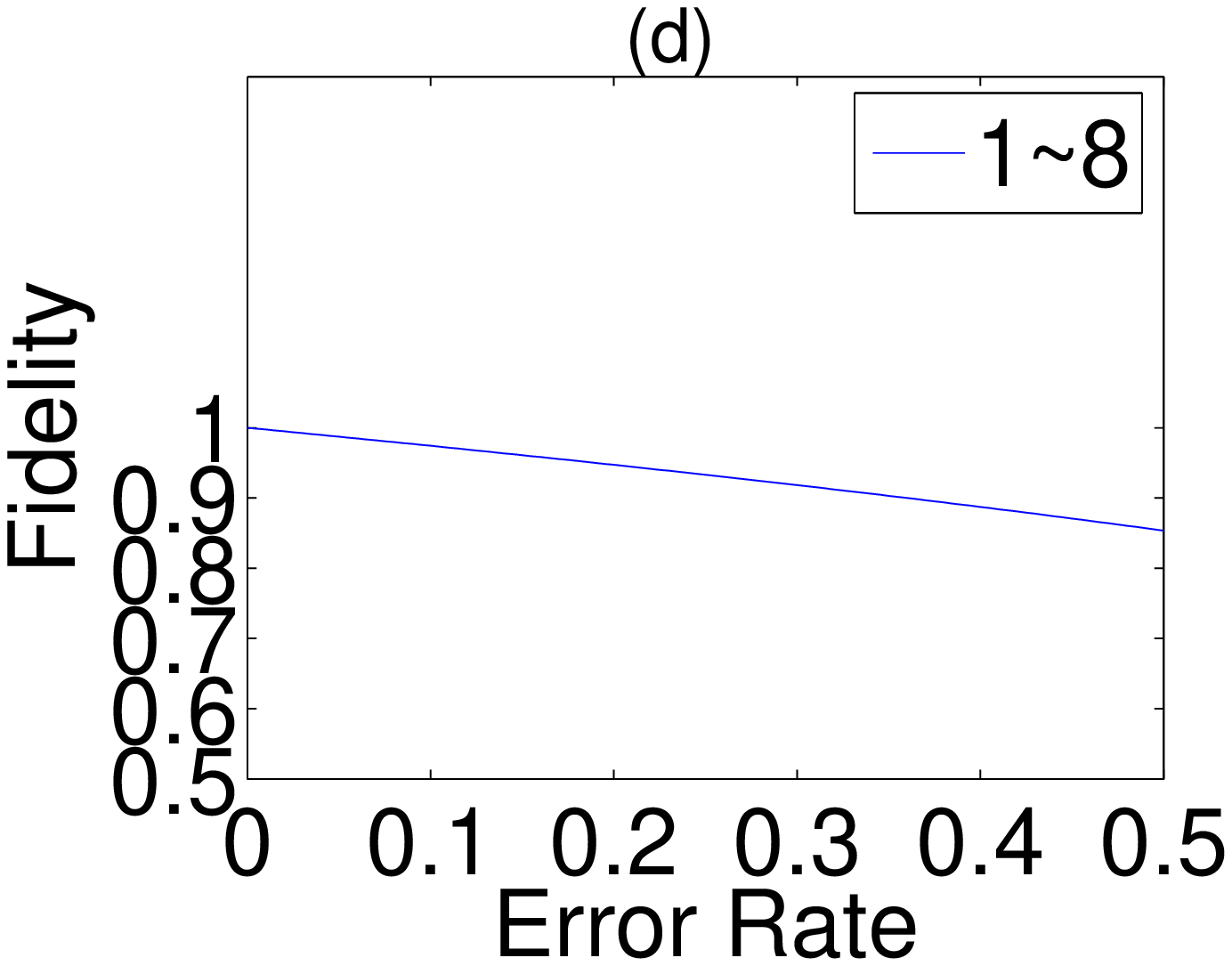}
\caption{The fidelity-error rate curve. Only one qubit is infected with Amplitude Damping noise, the numbers in the legend denote the qubit infected. Some of the results are the same and combined. (a)identity gate;(b)hadamard gate;(c)Z rotation gate;(d)controlled-Z gate; \label{Phasenoise}}
\end{center}
\end{figure}

(i) Vulnerability--For all those gates, the gate fidelity displays the vulnerability to both $X$ and $Z$ noises.
The fidelity decrease linearly with gradient -1 with the error probability $p$, reaching 0.5 at error rate 0.5.

The gates exhibit better resistance against amplitude damping channel and phase damping channel.
Particularly, for any gate, the gate fidelity holds above 0.85 at even error rate 0.5, which is still quit a good performance.

(ii) Immunity--As we see, there are several horizontal lines in the figures.
Interestingly, the gates are immune to certain types of error on certain qubits. This result seems counter-intuitive at the first sight, that if the qubit is exposed to certain type of noises, no damnify is taken on the implemented gate.
One can take advantage of this result in the process of constructing cluster state simply
by protecting a specified noise for certain qubits. Explicitly, if we know the main source
of noise, we just need to diminish the harmful one but not the noise which is actually immunized.

This phenomenon can be understood with the equation $P_xX\ket{\psi}=mP\ket{\psi}$, where $P$ is the projector of $X$.
We can understand that other operators such as $Z$ or $Y$ have similar phenomenon.
This equation demonstrates that the if the measurement base is coincident with the noise type, the effect of noise is equivalent to adding a global phase to the final gate we prepare, which does not damnify the accuracy of computation.
We remark that if a qubit is measured in $Z$ bases, the qubit is immune to phase damping noise too.

More interestingly, the immunity can be interpreted as freedom. Since the noise operator does not harm the result,
the initial states need not to be exact the needed cluster states.
To be explicit, to prepare a identical gate, besides the cluster states, the following 32 states as well as the mixed states of them are also optional:
\begin{eqnarray}
\rho_o=E^{(0)}E^{(2)}E^{(3)}E^{(4)}E^{(6)}\rho_{cl}E^{(6)\dagger}E^{(4)\dagger}E^{(3)\dagger}E^{(2)\dagger}E^{(0)\dagger}
\nonumber
\end{eqnarray}
where $E^{(0)},E^{(6)}=I~ \mathrm{or} ~Z,E^{(2)}, E^{(3)},E^{(4)}=I ~\mathrm{or}~ X$. This simplifies
the cluster states preparation.

\begin{figure}
\includegraphics[width=80mm]{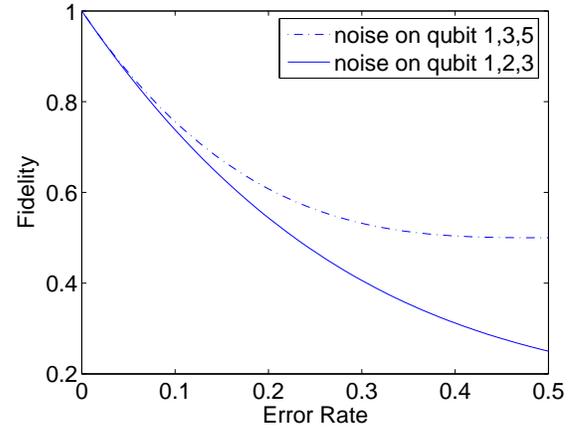}
\caption{A comparison of noise on qubits 1,3,5 and noise on qubits 1,2,3.   \label{controlpattern}}
\end{figure}

\section{Decoherence on multiqubit }

Next, we try to investigate a controlling pattern problem.
Suppose some number of qubits can be decoherence-free while others are exposed to inevitable decoherence,
how should we design the locations of those two different types of qubits.
We will show that there exists an effective controlling pattern which can improve the fidelity,
while there is also limitation of it.

We divide qubits of the initial cluster state into two groups--the controlled group and the exposed group. For simplicity, we assume that the qubits of controlled group is controlled perfectly, i.e., the qubit controlled has zero error rate.
The exposed group is in a noisy channel. We also assume that the error rate of the exposed group is the same. Such two assumptions are made to exclude other irrelevant elements and to highlight the point we find.

In the FIG \ref{controlpattern}, we show the numerical results of two controlling pattern. One curve denotes that we control the qubit 1,3 and 5 of the cluster state, the other one denotes that the we control the qubit 1,2 and 3. The fidelity of Hadamard gate is calculated and displayed in the FIG. \ref{controlpattern}.

FIG. \ref{controlpattern} illustrates that different controlling patterns leads to different fidelity behaviors.
By controlling every two qubits of original state£¬ one obtains better fidelity than by controlling the first three qubits.
This phenomenon can be briefly explained as follows. Recall the equation (\ref{a2}), the operator after $\rho$ is:
\begin{eqnarray}
\frac{1+\prod_{k=1}^{3} K_{2k-1}  }{2} \frac{1+\prod_{k=1}^{3} K_{2k}}{2}
\end{eqnarray}
As we see, the formula can not be written in the form--$F(Q_1)F(Q_2)F(Q_3)F(Q_4)F(Q_5)F(Q_6)$, where $F(Q_i)$ is a function of operators with only subscript $i$. Yet, this formula can be written in the form $F(Q_1,Q_3,Q_5)F(Q_2,Q_4,Q_6)$, thus the six qubits are divided into two groups--1,3,5 and 2,4,6. This means the six qubits do not 'stand lone', the every two qubits gathers in one group. Controlling the qubits of the same group leads to a better result than controlling the qubits of intergroup which can be shown by some calculations. This rule also goes for situation of two exposed qubits. E.g, allowing qubit 1,3 exposed to noise is a better choice than allowing qubit 1,2.

Despite the fact that we can choose a controlling pattern which may result in improvement
in fidelity of gates implementation, we also find the limitation of this technique.
We have the following observation: The derivative of the fidelity versus the error rate $p$
at the point when fidelity equals to one is independent of the controlling patterns.

Let us write the equation (\ref{a1})-(\ref{a4}) in the following form:
\begin{eqnarray}
F(\rho)=\tr (\rho_eM),
\end{eqnarray}
where
\begin{eqnarray}
\rho_e=\sum_{i_1,i_2\cdots i_k} E_{i_1}^{(j_1)}\cdots E_{i_{k}}^{(j_k)}\rho E_{i_k}^{(k)\dagger}\cdots E_{i_1}^{(1)\dagger}M,
\end{eqnarray}
here $M$ denotes operator that follows $\rho$ in equation (\ref{a1})-(\ref{a4}),
$k$ is the number of qubits in the exposed group. The superscript $j_1\cdots j_k$ labels the qubits in exposed group,
$i_1,i_2\cdots i_k=1,2$.

Next we calculate the $F$ derivation in terms of $p$. For any noise listed above,
we note that $E_2|_{p=0}=0$, and terms with multiplier $E_0$ equal to zero. We use superscript tildes to signify derivatives with respect to p. So, we have,
\begin{eqnarray}
\frac{\partial F}{\partial p}&=&\mathrm{Tr}[(2\tilde{E}_{2}^{(j_1)}E_{1}^{(j_2)}\cdots E_{1}^{(j_k)}\rho E_{1}^{(j_k)\dagger}\cdots E_{1}^{(j_2)\dagger}E_{2}^{(j_1)\dagger}\nonumber \\
&&+2E_{1}^{(j_1)}\tilde{E}_{2}^{(j_2)}\cdots E_{1}^{(j_k)}\rho E_{1}^{(j_k)\dagger}\cdots E_{2}^{(j_2)\dagger}E_{1}^{(j_1)\dagger}
\nonumber \\
&&\cdots \cdots \nonumber \\
&&+2E_{1}^{(j_1)}E_{1}^{(j_2)}\cdots \tilde{E}_{2}^{(j_k)}\rho E_{2}^{(j_k)\dagger}\cdots E_{1}^{(j_2)\dagger}E_{1}^{(j_1)\dagger}
\nonumber \\
&&+2\tilde{E}_{1}^{(j_1)}E_{1}^{(j_2)}\cdots E_{1}^{(j_k)}\rho E_{1}^{(j_k)\dagger}\cdots E_{1}^{(j_2)\dagger}E_{1}^{(j_1)\dagger}
\nonumber \\
&&+2E_{1}^{(j_1)}\tilde{E}_{1}^{(j_2)}\cdots E_{1}^{(j_k)}\rho E_{1}^{(j_k)\dagger}\cdots E_{1}^{(j_2)\dagger}E_{1}^{(j_1)\dagger}
\nonumber \\
&&\cdots \cdots \nonumber \\
&&+2E_{1}^{(j_1)}E_{1}^{(j_2)}\cdots \tilde{E}_{1}^{(j_k)}\rho E_{1}^{(j_k)\dagger}\cdots E_{1}^{(j_2)\dagger}E_{1}^{(j_1)\dagger})M]
\nonumber \\
\end{eqnarray}
Here we have used the identity $\rho^{\dagger}=\rho$, also for any noise listed above, $E_1|_{p=0}=I$.
Therefore the above equation can be rewritten in a more concise form:
\begin{equation}
\begin{array}{rl}
\frac{\partial F}{\partial p}=&\mathrm{Tr}[(2\tilde{E}_{2}^{(j_1)}\rho E_{2}^{(j_1)\dagger}+2\tilde{E}_{2}^{(j_2)}\rho E_{2}^{(j_2)\dagger} \\
&\cdots+\tilde{E}_{2}^{(j_k)}\rho E_{2}^{(j_k)\dagger} \\
&+2\tilde{E}_{1}^{(j_1)}\rho+2\tilde{E}_{1}^{(j_2)}\rho \\
&\cdots+2\tilde{E}_{1}^{(j_k)}\rho )M] \\
=&\mathrm{Tr}\{[(2\tilde{E}_{2}^{(j_1)}\rho E_{2}^{(j_1)\dagger}+2\tilde{E}_{1}^{(j_1)}\rho)\\
&+(2\tilde{E}_{2}^{(j_2)}\rho E_{2}^{(j_2)\dagger}+2\tilde{E}_{1}^{(j_2)}\rho)\\
&+\cdots +(\tilde{E}_{2}^{(j_k)}\rho E_{2}^{(j_k)\dagger}+2\tilde{E}_{1}^{(j_k)}\rho)]M\}\\
\end{array}
\end{equation}
This is exactly the summation of slopes of fidelities at point when each fidelity equals to one for noise occurring individually on qubits.
Thus we finish the calculation of our observation.

In the FIG \ref{controlpattern}, we can tell that the initial slope of both curve is 3 which is independent on controlling patterns. In fact, we know that the initial slope of any qubit 1,2,3,4,5 individually affected by noise $Z$ is 1.
Hence if $n$ qubits of qubits 1,2,3,4,5 are under individual noise, the initial slope of fidelity is $n$ independent on controlling patterns.

\section{Conclusion}

We have studied the success probability in implementing universal gates
of quantum computation under various noises in the well-known scheme of the
one-way measurement based quantum computation. We show that the locations of the qubits
in the cluster states play different roles for the fidelities of gate implementation.
In particular, some qubits are immune to a specified noise depending on
which measurement is performed.
This information can help us in designing efficient scheme of gates realization
for MBQC. Also we point out that we may take advantageous in choosing
appropriate controlling pattern for the cluster states, but this controlling pattern
does not help when the error rate is rather low. Those facts shed light on
the preparation of the cluster states for specified physical system.

Acknowledgements: This work was supported by 973 program
(2010CB922904), NSFC (11175248), NFFTBS (J1030310,J1103205), the Undergraduate Research Fund Of Education Fundation of Peking University, and
grants from Chinese Academy of Sciences.

\end{document}